\documentclass{article}
\usepackage[T1]{fontenc} 
\usepackage[utf8]{inputenc} 
\usepackage{ismir,amsmath,cite,url}
\usepackage{graphicx}
\usepackage{color}
\usepackage{xcolor}
\usepackage{arydshln}
\usepackage{soul}
\usepackage[bookmarks=false,hidelinks]{hyperref}
\usepackage{booktabs}
\usepackage{multirow}
\usepackage{svg}
\usepackage{subcaption}

\usepackage{lineno}

\newcommand{\eg}{e.\,g., }
\newcommand{\ie}{i.\,e., }
\newcommand{\wrt}{with reference to }

\newcommand{\sagg}{\ensuremath{S_\text{agg}}}
\newcommand{\nn}{\emph{RubricNet}}
\newcommand{\us}[1]{#1}

\title{Towards Explainable and Interpretable Musical \\Difficulty Estimation: A parameter-efficient approach}

\multauthor
{Pedro Ramoneda$^1$ \hspace{1cm} Vsevolod Eremenko$^1$ \hspace{1cm} Alexandre D'Hooge $^2$} { \bfseries{Emilia Parada-Cabaleiro$^3$ \hspace{1cm} Xavier Serra$^1$}\\
$^1$ Music Technology Group, Universitat Pompeu Fabra, Barcelona, Spain\\
$^2$  Univ. Lille, CNRS, Centrale Lille, UMR 9189 CRIStAL, F-59000 Lille, France\\
$^3$ Department of Music Pedagogy, Nuremberg University of Music, Germany\\
{\tt\small pedro.ramoneda@upf.edu}
\vspace{-0.7cm}
}

\def\authorname{P. Ramoneda, V. Eremenko, A. D'Hooge, E. Parada-Cabaleiro, X. Serra}

\begin{document}

\maketitle
\begin{abstract}
Estimating music piece difficulty is important for \us{organizing} educational music collections. This process could be partially \us{automatized} to facilitate the educator's role. Nevertheless, the decisions performed by prevalent deep-learning models are hardly understandable, which may impair the acceptance of such a technology in music education curricula. Our work employs explainable descriptors for difficulty estimation in  symbolic music representations. Furthermore, through a novel parameter-efficient white-box model, we outperform previous efforts while delivering interpretable results. These comprehensible outcomes emulate the functionality of a rubric, a tool widely used in music education.
Our approach, evaluated in piano repertoire \us{categorized} in 9 classes, achieved  
\(41.4\%\) accuracy independently, with a mean squared error (MSE) of \(1.7\), showing precise difficulty estimation. 
Through our baseline, we
illustrate how building on top of past research can  offer alternatives for music difficulty assessment which are explainable and interpretable. With this, we aim to promote a more effective communication between the Music Information Retrieval (MIR) community and the music education one.

\end{abstract}

\vspace{-0.2cm}
\section{Introduction}\label{sec:introduction}

Estimating the difficulty of music pieces aids in 
\us{organizing} large collections for music education purposes. 
However, manually assigning difficulty levels is laborious and might lead to
subjective errors~\cite{deconto2023automatic}. To address this, 
Music Information Retrieval (MIR) research has focused on automating this process for piano works represented in 
various modalities~\cite{ramoneda2022,ramoneda2024,zhang2023symbolic,ramoneda2023ismir,ramoneda2024taslp} as well as repertoires from 
other instruments~\cite{vasquez2023quantifying,holder2015musiplectics}. Furthermore, the interest of companies like Muse Group~\cite{musescore,ultimate} and Yousician~\cite{kaipainen2017system} highlights the industry's recognition of the importance of the task.

Previous work in this field has mainly focused on processing machine-readable symbolic  scores \cite{deconto2023automatic, sebastien2012score,chiu2012study,nakamura2014merged,nakamura2015automatic,ramoneda2022,ramoneda2024,zhang2023symbolic}.  
These, unlike acoustic features extracted from audio whose understanding depends on signal processing knowledge, are both \us{analyzable} by computers and interpretable by humans. Musicians find also easier to understand symbolic features since 
based on music theory knowledge. Initial 
works towards interpretable difficulty assessment focused on \us{visualization} 
\cite{sebastien2012score}, with Chiu and Chen
~\cite{chiu2012study} making the first attempt to classify difficulty in the piano repertoire with 
explainable descriptors. 
%
%
Still, the continually increasing trend towards deep-learning based solutions~\cite{zhang2023symbolic,ramoneda2024}, whose lack of transparency  limits  users'  understanding and therefore leads to an  eventual non-acceptance  in real life applications \cite{branley2020user}, can impair a fruitful implementation of such technologies in  music educational  practices.
\begin{figure}
    \centering
    \includegraphics[trim={0.0cm  0.0cm 0cm 0cm},clip,width=0.85\linewidth]{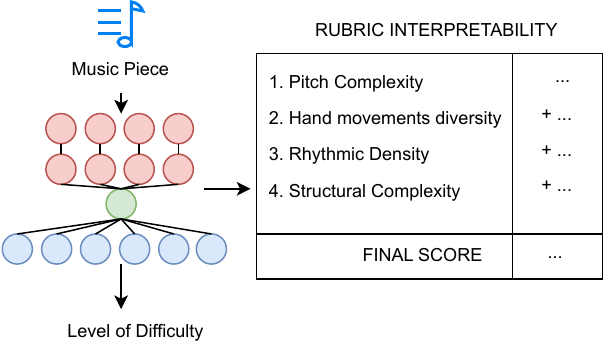}
    \vspace{-0.40cm}
    \caption{To promote a more objective and transparent assessment, in our  white-box model \nn, similarly as  educational rubrics,  scores (here difficulty) are dependent on  descriptors' values. The Rubric Interpretability table displayed at the right is inspired by~\cite[Fig.\ 1]{ustun2019jmlr}
    }
    \label{fig:teaser}
\vspace{-0.25cm}
\end{figure}

With this background, we propose a white-box~\cite{loyola2019black} model (cf.\ Figure~\ref{fig:teaser}), 
which 
through the concept of a 
rubric, \ie 
an evaluation instrument from 
music education  used  to support objective assessment ~\cite{Latimer2010Reliability,alvarez2020On,wesolowski2012understanding,jonsson2007the},  allows a transparent interpretation of music difficulty. From this point forward, the white-box model will be denoted by \nn. 
Furthermore, to gain a profound understanding of what 
music difficulty means from an explainable perspective, we build upon the 
descriptors  of  
Chiu and Chen
~\cite{chiu2012study} by proposing a new one focusing on music repetitive patterns. 
%
We also provide
an interactive companion page\footnote{At: \url{https://pramoneda.github.io/rubricnet}} to \us{visualize} 
the evaluated 
data and 
\us{scrutinize} the results 
in light of its 
interpretability from a musical point of view. 

\begin{figure}[t!]
    \centering
    \includegraphics[width=1\linewidth]{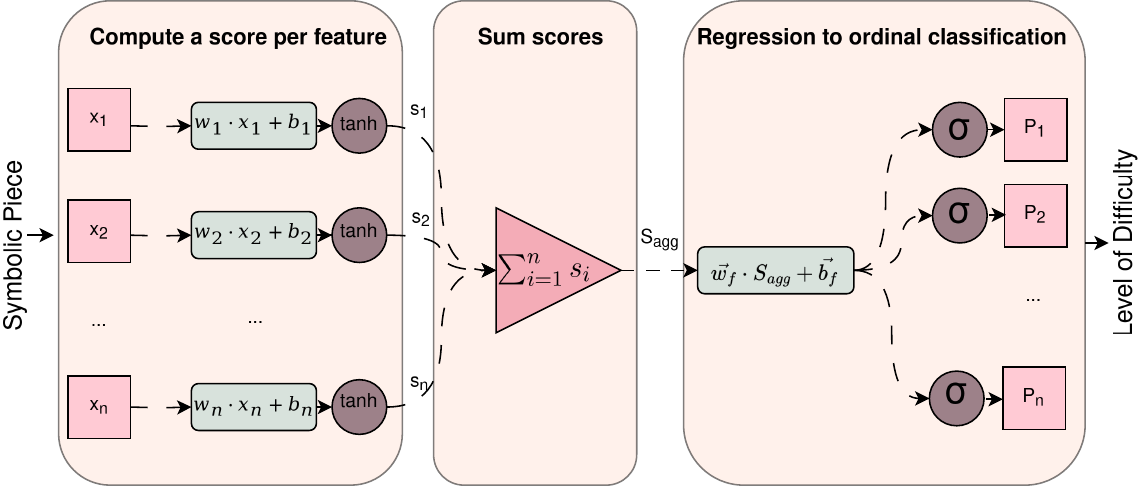}
    \vspace{-0.65cm}
    \caption{Detailed \nn's architecture.}
    \label{fig:rubricnet}
\vspace{-0.25cm}
\end{figure}

Through 
eXplainable Artificial Intelligence (XAI), we aim to contribute to music education by 
facilitating the understanding of measurable 
factors that determine a piece's difficulty. 
Our work builds on methods from 
music education, 
where 
objectively assessing  abstract competences through measuring concrete criteria 
is consolidated by employing rubrics, \ie tools which, unlike a black-box, break down complex concepts into simpler ones~\cite{Latimer2010Reliability,jonsson2007the}.

Our interpretable methodology aims to bridge the gap between computational models and practical music education needs, enabling educators to make facilitated, but also informed decisions about curriculum development based on the difficulty levels of pieces. 
We release all the code and models of this research\footnote{At: \url{https://github.com/pramoneda/rubricnet}}, in order to offer a baseline
for further research in music difficulty assessment.


\section{Related works}

Previous research aiming to  automatically assess  the difficulty of piano repertoire
examined the link between 
fingering patterns and the pieces'
difficulty level~\cite{nakamura2014merged,nakamura2015automatic,ramoneda2022}. %
Recent studies~\cite{alexandria,deconto2023automatic,ramoneda2024,zhang2023symbolic} have also made significant contributions. 
In \cite{ramoneda2024}, representations are used to feed three deep learning models—covering music notation, physical gestures, and expressiveness—to emulate Cook's dimensions~\cite{cook1999analysing}. These models' predictions are merged using an ensemble method to estimate the scores' difficulty. 
While we appreciate their musicology-inspired approach, its lack of interpretability harms its usability.
%

Difficulty estimation of 
piano pieces has also been investigated through 
hybrid methods that merge features with deep learning models~\cite{alexandria,deconto2023automatic}. However, the absence of publicly shared data and code complicates performing comparative analyses \wrt these works. 
In \cite{alexandria}, the authors  combine the methods from Chiu and Chen
~\cite{chiu2012study} with deep learning models trained using piano roll as input. 
In a similar vein, \cite{deconto2023automatic} 
uses JSymbolic features \cite{mckay2018jsymbolic} and deep learning models on a proprietary dataset.

In the study by Chiu and Chen
\cite{chiu2012study}, 159 pieces from the \textit{8notes} website were used, whereas \cite{alexandria} \us{utilized} 1800 MIDI files from the same source. The \us{categorization} of these pieces, provided by users of \textit{8notes}, raises concerns about their reliability. Unfortunately, neither study provides access to their data or details on how they were segmented. Other work \cite{morsi2024characterizing} has attempted to understand the effectiveness of various features, including those proposed in \cite{chiu2012study} for categorizing the grade levels of a specific piano curriculum. Recent efforts by Zhang et al. \cite{zhang2023symbolic} and Ramoneda et al. \cite{ramoneda2024} have focused on compiling datasets with difficulty annotations from the established piano publisher Henle Verlag, with the latter's dataset not only being the most extensive but also the only one made publicly available. Therefore, for our comparative analysis, we will use the  open-source datasets presented in \cite{ramoneda2024}, namely \emph{Can I Play It?} (CIPI), which has 9 levels of difficulty, and \emph{Mikrokosmos-difficulty} (MKD), which includes 3 levels.

Finally, in order to validate our approach, we consider a different and established feature set, i.e., the standard music symbolic features available through Music21 library~\cite{cuthbert2010music21}, which includes (amongst others) established JSymbolic features~\cite{mckay2018jsymbolic},  
thus facilitating a meaningful comparison with our proposed descriptors. In addition, 
we also contrast the results achieved with our novel descriptors with those obtained with the features by Chiu and Chen
~\cite{chiu2012study}, which are also reimplemented and open-sourced in this study.
Note that none
of the approaches previously mentioned 
has focused on the interpretability of the descriptors, which is a key contribution of our work.

\section{Interpretable \nn{}}

The \nn{} model (cf. Figure~\ref{fig:rubricnet}) is designed to provide interpretability akin to a rubric, enabling its analysis and results to be intuitively aligned with established practices in music education. This approach ensures that the model's logic and outcomes are easily comprehensible, facilitating their usage in music education 
along to traditional 
tools.

\subsection{Model Architecture}

The network, 
comprising a series of linear layers dedicated to process individual input descriptors and followed by a nonlinear activation function,  
is formulated as follows:

Given a set of $N$ input descriptors, each descriptor $x_i$ is first processed through its dedicated linear layer with weight 
$w_i$ and bias $b_i$, followed by a hyperbolic tangent activation function to yield:
\begin{equation}
s_i = \tanh(w_i \cdot x_i + b_i)
\end{equation}
where $s_i$ represents the processed score for the $i$-th descriptor. 
Scores are then aggregated  
in a single 
score $S_{agg}$:
\begin{equation}
S_{agg} = \sum_{i=1}^{n} s_i
\end{equation}
The aggregated score $S_{agg}$ is then 
passed through a final linear layer to obtain the logits for the class predictions, which are  
mapped to probabilities with a 
sigmoid function:
\begin{equation}
\vec{P} = \sigma(S_{agg} \cdot \overrightarrow{w_f}  + \overrightarrow{b_f})
\end{equation}
where $\sigma$ denotes the sigmoid function, 
$\overrightarrow{w_f}$ and $\overrightarrow{b_f}$ are the weight and bias of the final linear layer, respectively. 


\subsection{Ordinal \us{Optimization}}

This model applies an ordinal \us{optimization} approach~\cite{cheng2008neural}, predicting ordered categorical outcomes, \ie  difficulty levels such as beginner (1), intermediate (2), and  advanced (3), 
through logits. These logits, computed using a mean squared error (MSE) loss, indicate the model's predictions on the ordinal scale.
Difficulty level is then obtained as:
\begin{equation}
    \max \{ i \text{ where } P_i \geq 0.5 \text{ and } P_j \geq 0.5, \forall j < i\}
\end{equation}
\vspace{-0.8cm}

\subsection{Interpretability}

In \nn{}, the \textit{descriptors} (automatically computed from the data) are, to some extent, comparable to the \us{formalized} \textit{evaluation criteria} defined in traditional rubrics; similarly, the \textit{aggregated score}, might be comparable to a final \textit{grade/mark} assigned in an educational scenario. Given the correspondences between both, we could consider the model a ``white-box'' approach, able to promote 
transparency and interpretability, similarly to a  
rubric.

It uses independent linear transformations on input descriptors to generate scores between -1 and 1, which directly influence the regression output, $S_{agg}$. 
Since negative scores, might be not fully understood in terms of difficulty level, 
we \us{normalize} scores between 0 and 1, rescaling $\sagg$ between 0 and 12. 
This approach mirrors rubric's ability to provide objective and 
structured feedback, with the simplicity of these transformations aiding in understanding the impact of features in predictions.

The interpretability of the model lies in its ability to dissect each descriptors' influence on a piece's difficulty level. 
Consequently, \us{analyzing} each descriptor's scores might reveal its overall importance on the prediction.
Lastly, $S_{agg}$ is a continuous-ordered scalar with rank correlation to difficulty. 
Therefore, from $S_{agg}$, we retrieve ordered and discrete categories  with clear decision boundaries.

\section{Explainable Descriptors}

\begin{table}[t!]
\small
\centering
\begin{tabular}{p{0.25\columnwidth} p{0.65\columnwidth}}
\hline
\textbf{Descriptor} & \textbf{Explanation} \\
\hline
Pitch Entropy & Indicates pitch variety; higher values mean more diverse pitch collection\\
Pitch Range & 
Distance between the lowest and highest notes.  \\
Average Pitch & Indicating the central pitch level. \\
Displacement Rate & Measures hand movement intensity across keys reflecting physicality in performance. \\
Average IOI & The average timing between note onsets, indicative of rhythmic density. \\
Pitch Set LZ & Indicative of structural complexity and repetitiveness within a pitch set sequence. \\
\hline
\end{tabular}
\vspace{-0.25cm}
\caption{Explanation of descriptors in musical terms.
}
\label{tab:music_descriptors}
\vspace{-0.25cm}
\end{table}


From codified musical scores, we extracted numeric features which are feed to a classification
algorithm. 
We re-implemented a set of features from the literature~\cite{chiu2012study} while proposing
a novel one,
Pitch Set LZ.
In addition to explaining the features (cf. Table~\ref{tab:music_descriptors}), we will provide their technical descriptions and \us{analyze} their relevance to difficulty and interdependencies using the data.

\subsection{Descriptors}

In our work, we \us{analyze} music sheets encoded in symbolic format, focusing on extracting pitch and timing. Following the approach suggested by Chiu and Chen \cite{chiu2012study}, we process left and right hand parts separately to clarify pedagogical aspects of musical difficulty. Our primary analysis involves sequences of pitch set events, each \us{characterized} by a pitch set $S$ and onset time $T$. Pitch sets, represented by sets of MIDI numbers, are defined over the alphabet of all pitch sets $\mathbf{S}$ that occurred in a score part, while onset times are calculated in seconds from the performance start by the music21 library \cite{cuthbert2010music21} \wrt marked tempo information. This method \us{emphasizes} the timing of note attacks, duration and rests. Additionally, we consider a collection of pitch events, each defined by pitch $P$ over the alphabet of all pitches $\mathbf{P}$. Our analysis started with the five features identified by Chiu and Chen \cite{chiu2012study} as most relevant to understanding musical difficulty.

\vspace{0.15cm}
\noindent
\textbf{Pitch Entropy}.
The entropy of pitches in the pitch events:
\begin{equation}\label{pitchentropy}
-\sum_{i\in \mathbf{P}} p(P = i)\log_{2}p(P = i)\end{equation}

\noindent
\textbf{Pitch Range}.
The distance between the minimum and maximum MIDI pitches in a score part.

\vspace{0.15cm}
\noindent
\textbf{Average Pitch}.
The average MIDI pitch in a music sheet.

\vspace{0.15cm}
\noindent
\textbf{Displacement Rate}. 
Initially proposed by \cite{chiu2012study}, it quantifies the extent of hand movement across the
keyboard during the performance of a score. It \us{analyzes} maximum pitch distances between consecutive pitch set events and is calculated as a weighted average of three categories: distances less than 7 semitones (assigned a weight of zero); distances over 7 semitones but under an octave (assigned a weight of one); and distances of an octave or larger (assigned a weight of two to \us{emphasize} larger movements).

\vspace{0.15cm}
\noindent
\textbf{Average IOI}:
Average Inter Onset Interval.
A concept similar to the ``Playing speed''  
introduced by
~\cite{chiu2012study}, a term we consider deceptive since it actually decreases as the hand's ``speed'' increases. This is an average time in seconds between onsets of two consecutive pitch set events. Let's denote $i^{th}$ onset time with $T_i$, then the value is:
\begin{equation}\label{averageioi}
\frac{\sum_{1\le i \le N^{events} - 1}(T_{i+1} - T_i)}{N^{events}-1}
\end{equation}
In 23\% of the scores, information about the recommended performance tempo is missing.
We then assume the tempo is 100 beats per minute (bpm). 
Thus, in cases of missing bpm, the Average IOI feature might not be relevant.

\vspace{0.15cm}
\noindent
\textbf{Pitch Set LZ}. Lempel-Ziv complexity of pitch set sequence. Before introducing our proposed descriptor,
it is crucial to provide context and motivation.
Pitch Entropy, as \us{emphasized} by Chiu and Chen \cite{chiu2012study},
is particularly relevant---a conclusion supported by the analysis of correlations between difficulty
and features in the following section, as well as by informal experiments.
As Sayood discusses~\cite{Sayood2018}, there's a link between entropy of a task
and the cognitive load it imposes on the performer, a concept that may also apply to music performance \cite{Palmer2012}.
However, music is often perceived in terms of larger structures like phrases
and sections, not just isolated pitches, prompting us to seek a descriptor that captures the ``repetitiveness''
of music on a broader scale. To this end, we employ LZ-complexity, a measure of redundancy introduced by Lempel and Ziv \cite{ziv1978compression}.
In context of music research, it was used for binary encoded rhythm analysis by Shmulevich and Povel
\cite{shmulevich2000measures}. We apply LZ-complexity to sequence of pitch sets:
scan a score part, identify all subsequences of pitch sets that cannot be reproduced
from preceding material through a recursive copying procedure.
The number of such unique subsequences is defined as the LZ-complexity of the part.
This approach allows us to assess the structural complexity and redundancy of a musical piece,
highlighting the cognitive demands placed on performers.

\subsection{Feature Analysis}
\label{sec:feature_analysis}

\begin{table}[t!]
\centering
\small
\begin{tabular*}{\columnwidth}{l@{\extracolsep{\fill}}r}
\hline
\textbf{Feature} & \textbf{$\tau_c$} \\
\hline
Pitch Entropy (R) & 0.583 \\
Pitch Set LZ (L) & 0.583 \\
Pitch Entropy (L) & 0.582 \\
Pitch Set LZ (R) & 0.573 \\
Pitch Range (L) & 0.567 \\
Pitch Range (R) & 0.554 \\
Displacement Rate (R) & 0.332 \\
Displacement Rate (L) & 0.273 \\
Average IOI (R) & -0.209 \\
Average IOI (L) & -0.208 \\
Average Pitch (R) & 0.088 \\
Average Pitch (L) & 0.017 \\
\hline
\end{tabular*}
\vspace{-0.25cm}
\caption{Features ordered by absolute values of their $\tau_c$ rank correlation with the difficulty level.
}
\label{tab:tau_c}
\vspace{-0.25cm}
\end{table}

We assume that, for easier interpretability,
features must on average change monotonically with the difficulty level.
To measure this quality, 
we use the $\tau_c$ version of Kendall rank correlation coefficient due to 
its ability to deal with
``heavily tied'' rankings \cite{Kendall1962} (many musical pieces have the same difficulty, hence, we have multiple ties in the ranking by difficulty). 
$\tau_c$ is equal to 1 when feature and difficulty rankings are perfectly aligned in the same direction, -1 if they are aligned in opposite directions. As the number of nonconcordant cases increases, the coefficient approaches zero. 
In \autoref{tab:tau_c},
the results show that
the features related to pitch \us{organization} are the most correlated to difficulty.
Hand displacement and Inter-onset intervals are less correlated,
while average pitch seems almost irrelevant.

\begin{figure}[t!]
    \centering
    \includegraphics[trim={0.0cm  0.25cm 0cm 0cm},clip,width=\columnwidth]{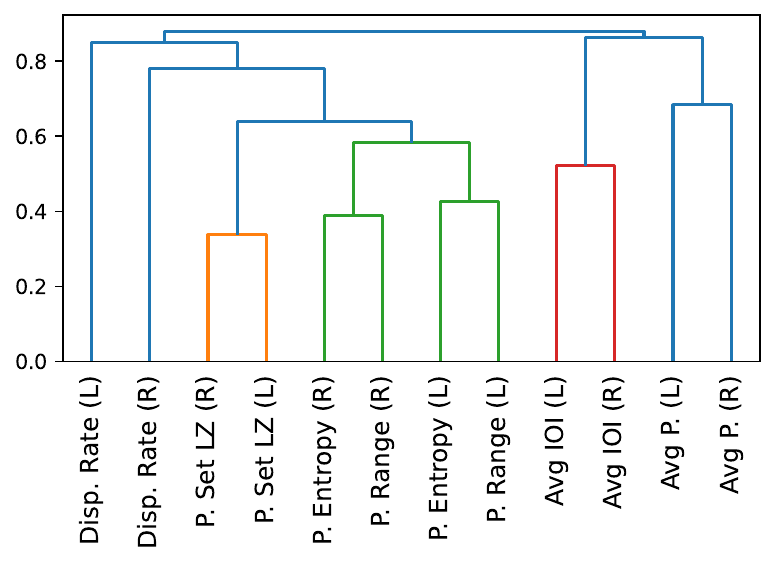}
\vspace{-0.65cm} 
    \caption{Hierarchical clustering of features based on their average correlation distance within each difficulty class. 
    }
    \label{fig:dendro}
\vspace{-0.25cm}
\end{figure}

In addition, we aim to uncover dependencies among the features themselves while mitigating the influence of difficulty, with whom  most features are correlated.
To achieve this,
we calculate conditional $\tau_c$ correlations for all feature pairs given a fixed difficulty level, 
and average the coefficients across all difficulty levels.
We then convert these coefficients into a distance matrix
and apply hierarchical agglomerative clustering based on average distance to identify clusters of correlated features.
From the resulting dendrogram (cf. \autoref{fig:dendro}), we observe that features correlated with
difficulty---namely Pitch Entropy, Pitch Set LZ, and Pitch Range---are also interrelated.
This is remarkable because the three most correlated features are not inherently dependent:
 one could envision a music piece with any of them \us{maximized} while maintaining low values for the others.
However, pieces in CIPI
typically exhibit coordinated values in these descriptors.
Thus, we mostly observe the combined effect of these features,
making it challenging to reliably decompose ``difficulty'' into an aggregate of independent components.

\vspace{-0.25cm}

\section{Experiments}
\subsection{Experimental Setup}

To evaluate the effectiveness of our proposed method, we \us{utilized} the \emph{Mikrokosmos-difficulty} (MKD), and \emph{Can I play it?} (CIPI) datasets~\cite{ramoneda2024}.
For fair comparison,
we use the 5-fold cross-validation approach defined in \cite{ramoneda2024}. 
In each split, 60\% of the data is used as a train set, while the remaining  is equally divided into  validation and test sets.

As in~\cite{ramoneda2024},  
we employ 
mean squared error (MSE) and accuracy within $n$ classes (\(\mbox{Acc}\text{-}n\)) for evaluation. These metrics are chosen for their applicability to ordinal classification challenges, with \(\mbox{Acc}\text{-}n\) assessing the model's accuracy for $n$ classes from the true labels, and MSE measuring the average squared prediction error across classes.
The effects of dataset imbalances and a fair evaluation across classes are mitigated by macro-averaged metrics.

We \us{optimize} the models during training through Adam \us{optimizer} with a learning rate of \(10^{-2}\).
The training process incorporates early stopping, based on the \(\mbox{Acc}\text{-}n\) and MSE metrics from the validation set, to prevent overfitting. Through 
Ordinal Loss, 
we frame difficulty prediction as an ordinal classification task, as mentioned in Section 3. 
We apply a standard scaler and dropout to the features to prevent individual ones 
from dominating. 
For each experiment, we look for the best hyperparameters using Bayesian \us{optimization}~\cite{akiba2019optuna}: batch size within the range from 16 to 128, dropout rate between 0.1 and 0.5, learning rate decay from 0.1 to 0.9, and the learning rate itself, tested over a logarithmic scale from \(1e-5\) to \(1e-1\). This 
approach allows us to systematically explore the hyperparameter space and identify the optimal settings for our models; thus, enabling a
fair comparison between experiments.


\subsection{Experimental Results}

\begin{table}[t!]
\small
\centering
\begin{tabular}{p{0.30\columnwidth}p{0.16\columnwidth}p{0.16\columnwidth}:p{0.16\columnwidth}}
\cline{2-4}
          & \multicolumn{2}{c:}{CIPI}                            & \multicolumn{1}{c}{MKD}    \\ \cline{2-4} 
          & \multicolumn{1}{c}{Acc-9}                   & \multicolumn{1}{c}{MSE}        & \multicolumn{1}{c}{Acc-3}   \\ \hline
argnn~\cite{ramoneda2024}    & 32.6(2.8)  & 2.1(0.2) & 75.3(6.1)  \\
virtuoso~\cite{ramoneda2024} & 35.2(7.3) & 2.1(0.2) & 65.7(7.8)  \\
pitch~\cite{ramoneda2024}        & 32.2(5.9)   & 1.9(0.2) & 74.2(9.2) \\
ensemble~\cite{ramoneda2024} & 39.5(3.4) & \textbf{1.1(0.2)} &  76.4(2.3) \\
\hdashline
\textbf{Ours} & \textbf{41.4(3.1)} & 1.7(0.5) & \textbf{79.6(8.8)} \\
\hline
\end{tabular}
\vspace{-0.25cm}
\caption{Experiment comparison of previous individual deep learning models~\cite{ramoneda2024}, their ensemble and our explainable and interpretable method on CIPI and MKD.}
\label{tab:final_cipi}
\vspace{-0.15cm}
\end{table}

In Table~\ref{tab:final_cipi}, the results from the comparison between the performance of our novel approach  with the presented descriptors  (cf.\ Sections 3 and 4) and the results achieved by three previous models from the literature (argnn~\cite{ramoneda2022automatic}, virtuoso~\cite{jeong2019virtuosonet}, pitch) as well as their collective ensemble, are shown. 
Our model 
achieves the highest
Acc-9 score of $41.4 (\pm3.1)$ in CIPI, surpassing the ensemble's $39.5 (\pm3.4)$, 
while displaying the second lower 
MSE of $1.7 (\pm0.5)$, only overtaken by 
the ensemble's $1.1 (\pm0.2)$. 
With an Acc-3 score of $79.6 (\pm8.8)$ in the MKD dataset, our approach is superior to previous ones but with a higher standard deviation.

In the following, we 
examine the impact of various feature and model configurations on \nn{} performance (cf.\  Table~\ref{tab:ablation}).
As baseline for comparison, we consider the configuration previously discussed (cf.\ Ours in Table~\ref{tab:final_cipi}).

Employing only the five Chiu and Chen~\cite{chiu2012study} descriptors, \ie excluding Pitch Set LZ, leads to a decrease in Acc-9 by $-5.2$, reflecting a performance drop from the baseline. The use of Music21~\cite{cuthbert2010music21} descriptors, which include JSymbolic~\cite{mckay2018jsymbolic} and other descriptors widely used in the community, results in a decrease in Acc-9 by $-4.7$ and a decrease in MSE by $-0.4$, showing slight improvements in MSE but not in accuracy. However, note that a larger number of descriptors could decrease the explainability.
Combining all the descriptors  
slightly decreases Acc-9 and MSE:   $-2.5$ and $-0.4$, respectively; with the accuracy results  still under the baseline. 
These results indicate that 
the  descriptors discussed 
in Section 4 constitute the best option for difficulty estimation on CIPI. 
Since average pitch showed no relation to difficulty in the feature analysis, we repeated the experiments without this feature. This lead, however, to non-significant worsening of the results.

\begin{table}[t!]
\centering
\small
\begin{tabular*}{\columnwidth}{lll}
\hline
Experiment          & Acc-9           &   MSE  \\ \hline               
\nn{} proposed& \textbf{41.4(3.1)} & 1.7(0.5)\\ 
"" with Chiu and Chen~\cite{chiu2012study} descriptors& 36.2(5.2)  & 1.7(0.3)\\  
"" with Music21 descriptors & 36.7(6.0)  & 1.3(0.2)\\ 
"" with ALL descriptors & 38.9(4.3)  & \textbf{1.3(0.1)}\\ 
"" proposed without Avg P. & 39.0(5.6) & 1.5(0.4)\\
\hdashline 
"" with positive scores&38.5(3.5)  & 1.6(0.6)\\  
"" without ordinal regression & 36.2(1.3)  & 2.1(0.4)\\  
Logistic regression& 40.0(4.3)  & 1.5(0.3)\\ \hline 
\end{tabular*}
\vspace{-0.25cm}
\caption{Ablation study results for  different feature sets (5 first rows) and model configurations (last 3 rows) on  CIPI.}
\label{tab:ablation}
\vspace{-0.45cm}
\end{table}

Concerning the impact of different model configurations, we replace the tanh by sigmoid non-linearities to guarantee positive scores. The obtained MSE rate is similar but the accuracy drops by $-3.1$. This means that negative scores could aid in training, which is why we keep them, but \us{normalize} the scores after the training phase. Besides, substituting the ordinal encoding used in the baseline 
with a traditional one-hot encoding with cross-entropy loss, results in a decrease in Acc-9 by $-5.2$ and an increase in MSE by $+0.4$, highlighting the importance of ordinal regression in achieving lower MSE rates. Lastly, logistic regression with ordinal loss 
decreases the Acc-9 by $-1.4$ while showing a decrease of MSE by $-0.2$. This offers a compromise for both metrics
but without beating our setup and to our understanding, being less interpretable.  

Overall, the gains offered by Rubricnet with the features proposed are relatively modest compared to the baselines. However, having a 
smaller feature set is necessary for explainability. The novelty of our approach lies in aligning the interpretability of music education with rubric-like interpretability feedback. This alignment is essential for a successful 
application 
of our model in 
practice, as we will discuss in further sections. 

\subsection{Decision Boundaries}

\begin{figure}[t!]
    \centering
    \includegraphics[trim={0.0cm  1.0cm 0cm 0.0cm},clip,width=\columnwidth]{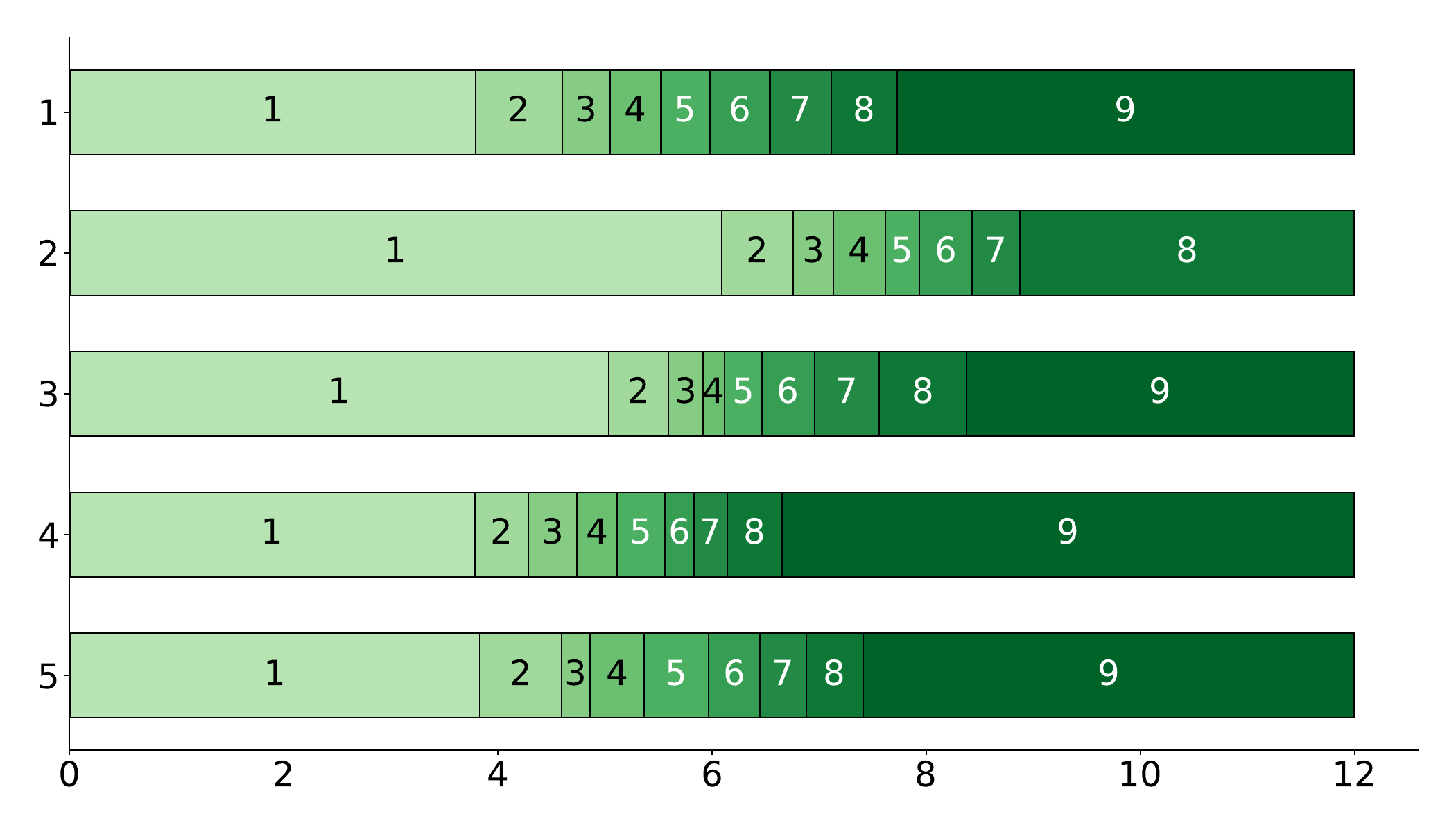}
\vspace{-0.65cm}
    \caption{Decision boundaries of the model between grades on $\sagg$ ($X$ axis) for all splits ($Y$ axis) on CIPI.} 
    \label{fig:boundaries}
\vspace{-0.25cm}
\end{figure}

In \nn{}, the input features are combined into a single scalar before performing
the final ordinal classification. Analysis of the results shows that the final layer
defines \us{optimized} decision boundaries, setting thresholds for $S_\text{agg}$ that progressively increase along with difficulty levels.
Because of the final sigmoid activation, once $\sagg$ exceeds a boundary, the 
corresponding difficulty level will always be active, which guarantees the ordinality of the predictions.
%
By examining the decision boundaries (cf.\ Figure~\ref{fig:boundaries}), we observe that the trends are similar across splits, displaying shorter valid ranges around intermediate levels. Note that in split 2, there are only 8 classes because the model ignored the last class. This can happen as we use numeric optimization, which sometimes falls into local minima. These minima might seem optimal based on the validation metrics but do not meet our overall performance expectations.

\section{Discussion and limitations}

\begin{figure}[t!]
    \centering
    \includegraphics[trim={0.0cm  1.1cm 0cm 0.7cm},clip,width=\columnwidth]{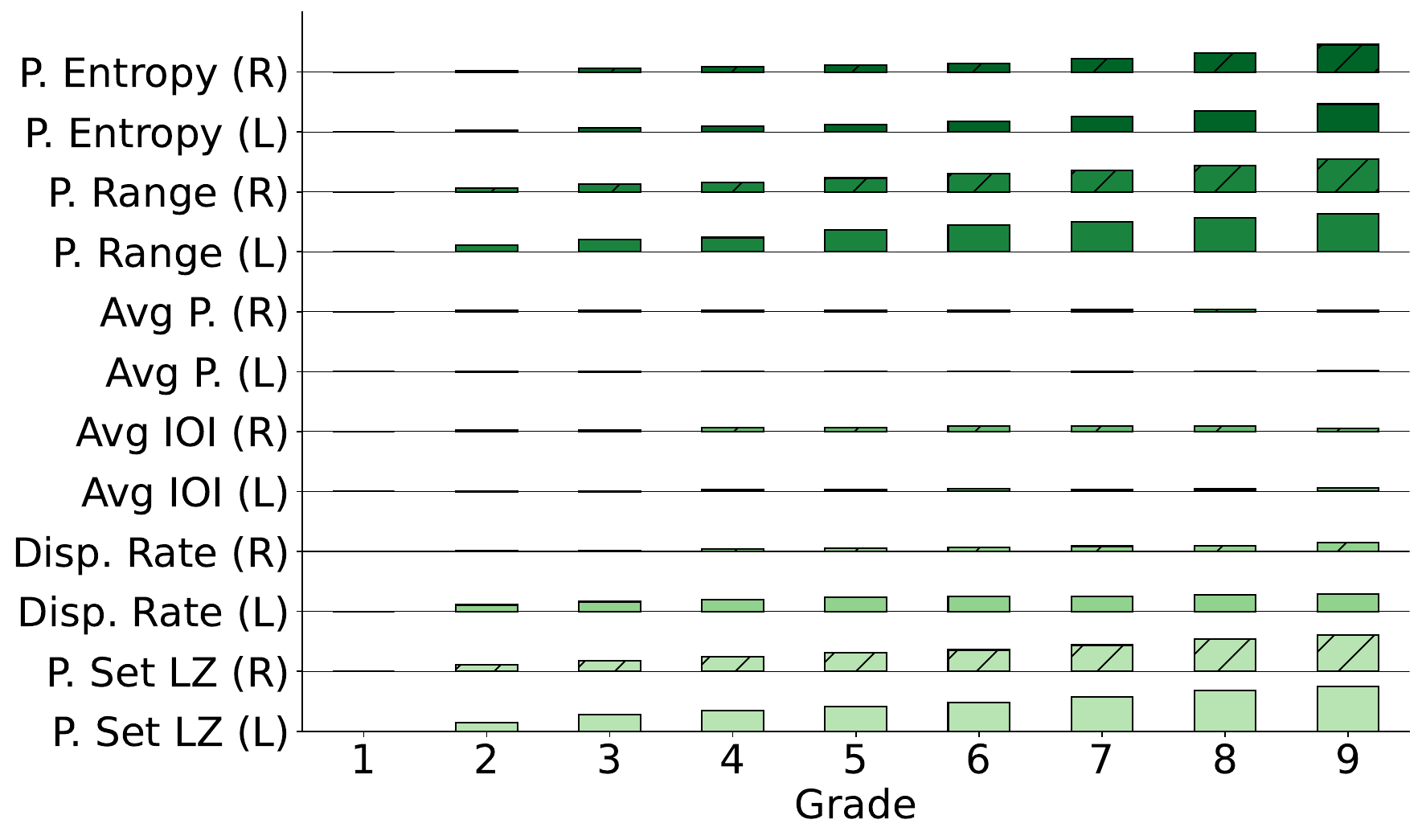}
    \vspace{-0.75cm}
    \caption{Average relative contribution 
    of  descriptors ($Y$ axis)  normalized between 0 and 1, across grades ($X$ axis).} 
    \label{fig:feat-contrib}
\end{figure}

Now, we \us{analyze} whether \nn{} is interpretable from a musical point of view. 
To understand  how features impact the final level suggested by the model, we evaluate the contribution of each descriptor to the aggregated score. 
Since learning to play an instrument is a progressive process,  relative contributions of features to different levels \wrt the grade 1 are displayed instead of  absolute values.
These contributions are averaged across splits on the test set and shown in  \autoref{fig:feat-contrib}.

\begin{figure}[t!]
    \centering
    \includegraphics[trim={0cm  0cm 0cm 0cm},clip,width=0.96\linewidth]{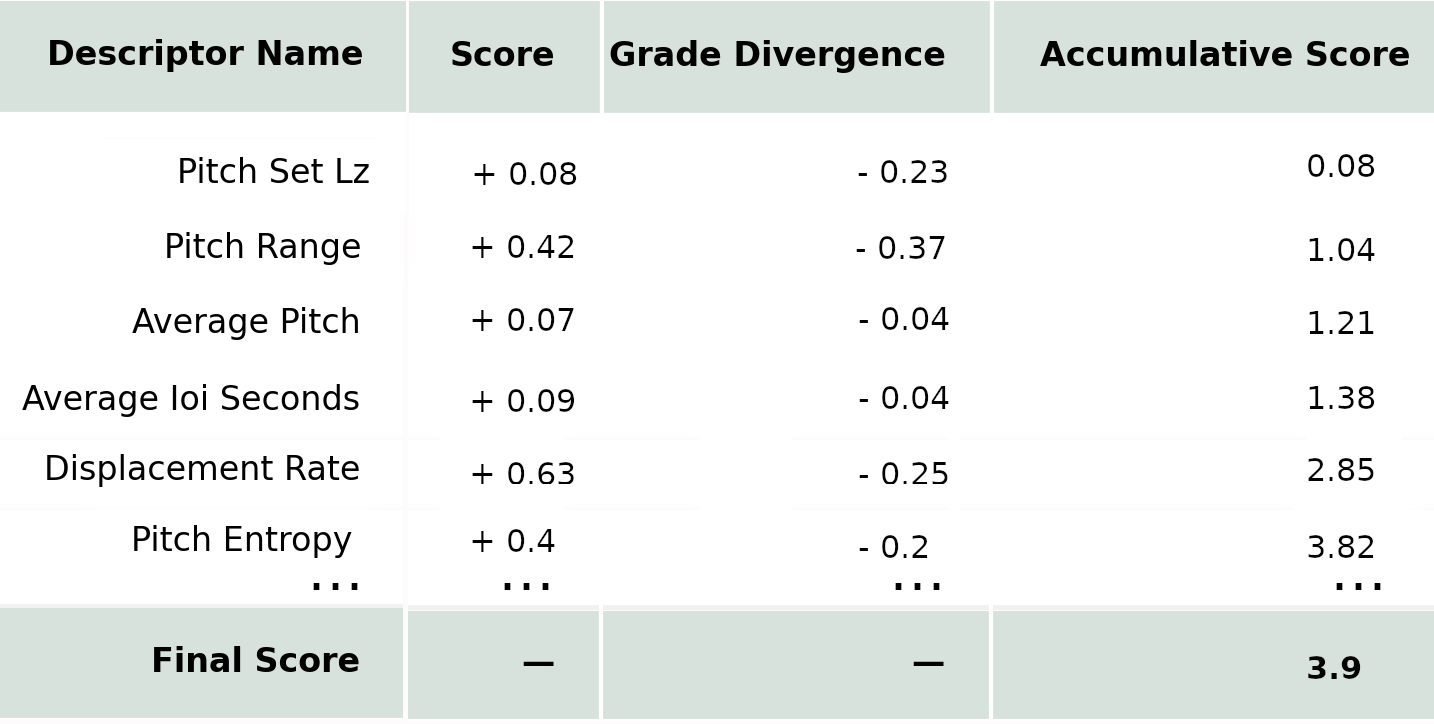}
\vspace{-0.25cm}
    \caption{Simplified 
    difficulty  interpretable rubric for the  \emph{Nocturne op.\,9, no.\,3}  (F. Chopin). Descriptors' values for the right hand   and  final score (for both hands) are shown. 
    }
    \label{fig:interpretable-nocturne}
\vspace{-0.35cm}
\end{figure}

We observe a trend of higher contributions when the level
increases for every descriptor. 
This observation is consistent 
with the fact that $\sagg$ value increases for higher levels (cf.\ \autoref{fig:boundaries}).
The most discriminative features are 
pitch entropy and 
pitch range, 
as well as the LZ descriptor for 
higher levels. 
Conversely, some features, \eg average IOI 
or the average pitch,  
have low contribution 
to the model's decisions, as shown by their relatively constant and small values across grades. 
The latter is expected, since 
very different pieces could have the same average pitch, not disclosing anything about difficulty.
The former might be explained by the averaging, which
can remove information, especially when a piece can alternate between fast and slow parts. Besides, as mentioned before, tempo is often poorly annotated in the dataset.


To better understand the explainable capabilities of the proposed descriptors, in the following, we provide a musical examination of two concrete samples, by this demonstrating the interpretability of our approach.
\emph{Nocturne op.\,9, no.\,3} by F. Chopin, is labelled as level 7, but classified as level 2. All the descriptors are below the grade average, as shown in the rubric (cf.\ \textit{Grade Divergence} in Figure~\ref{fig:interpretable-nocturne}).
Our hypothesis is that this nocturne contains many challenges that go beyond the descriptors used. There are constant changes in dynamics, a variety of articulations, and as a key difficulty aspect, many types of polyrhythms between the right and left hands. Further research should address all the types of difficulty challenges, probably underrepresented in the existing datasets.

\begin{figure}[t!]
    \centering
    \begin{subfigure}[b]{0.95\linewidth}
        \includegraphics[trim={0.0cm  0.0cm 0cm 10.5cm},clip,width=\linewidth]{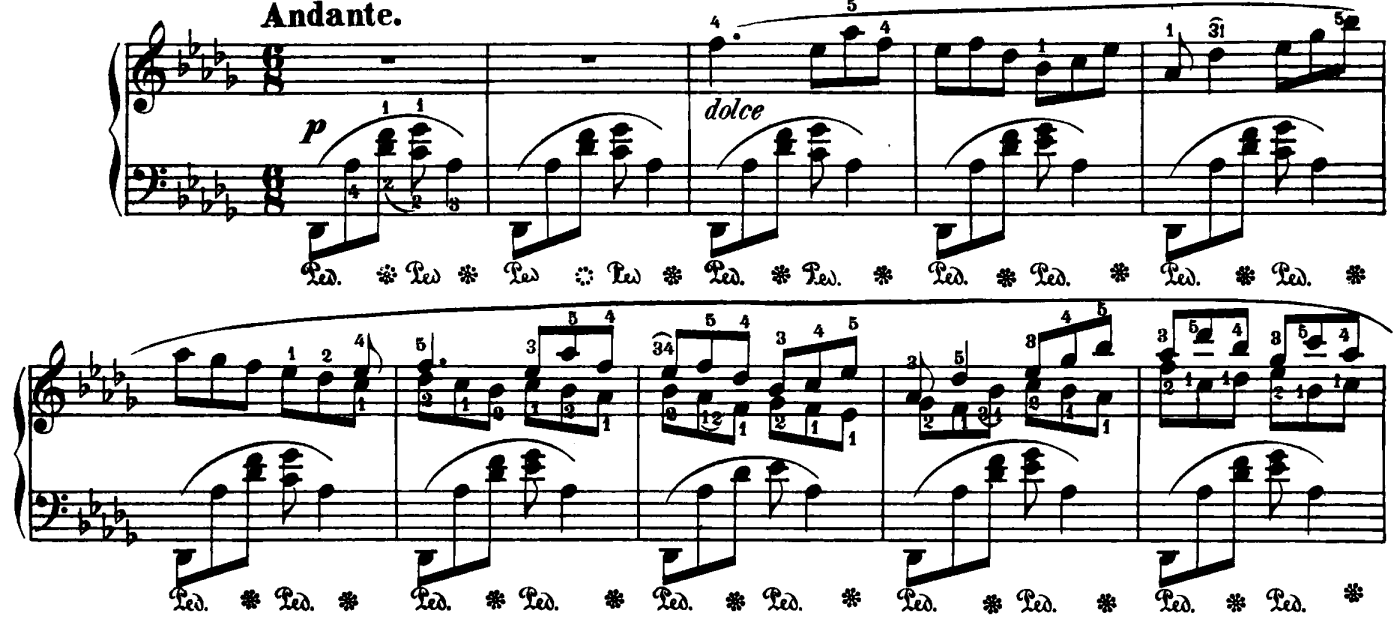}
        \caption{\textit{Berceuse in D-flat major, Op.57} (F.\ Chopin).  Bars 6-10.}
        \label{fig:chopin-excerpt}
    \end{subfigure}
    \begin{subfigure}[b]{0.99\linewidth}
        \includegraphics[trim={0.5cm  0.5cm 0cm 0.0cm},clip,width=\linewidth]{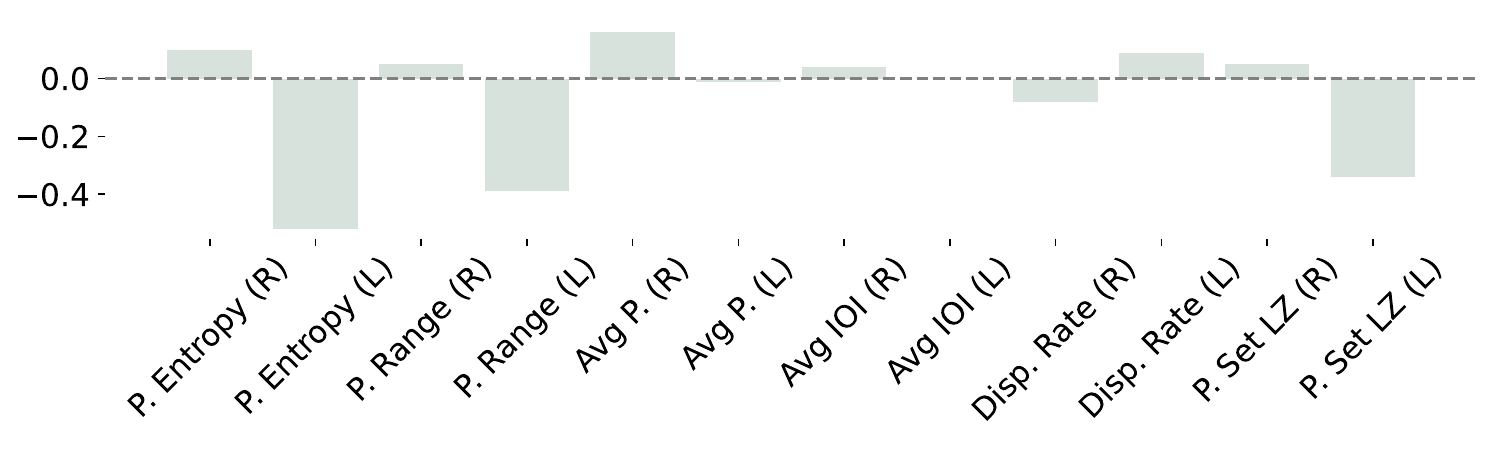}
        \caption{Distance to the average for the grade of the scores. Extracted from original rubric (grade divergence column). }
        \label{fig:chopin-scores}
    \end{subfigure}
    \vspace{-0.25cm}
    \caption{Musical excerpt (a) and a rubric outcome (grade divergence) plotted (b) from a piece in level 7.
    }
    \label{fig:chopin}
\vspace{-0.25cm}
\end{figure}

The piece \textit{Berceuse in D-flat major, Op.57} by F. Chopin, shown in \autoref{fig:chopin} is appropriately classified as grade 7. This is because it maintains a left-hand accompaniment with few changes, in contrast to the higher virtuosity of the right hand. 
The left hand has scores below average in most descriptors because of its few changes: Pitch Range (-0.41), Average IOI (-0.04), Pitch Set LZ (-0.34), Average Pitch (-0.01), and Pitch Entropy (-0.52). In contrast, the right hand shows more virtuosity, with higher than average scores for all the right hand features.
These scores collectively contribute to a final cumulative score that accurately reflects the overall difficulty.

Finally, it should  be noted that our approach primarily focuses on
descriptors
related to pitch sequences and
onsets, while disregarding others. 
Still, the ablation study showed that other features sets (\eg those from music21), even  covering aspects like rhythm variety,  
do not  enhance our classifier's performance either. In addition, expressive elements \cite{zhang2023disentangling} such as dynamics, tempo changes, and articulation, since often left to performers' interpretation, 
are  not always captured in musical notation \cite{ramoneda2024}, 
and therefore is 
a dimension our score-based model does not consider.

\vspace{-0.2cm}
\section{Conclusion and future work}

In our study, we proposed a novel white-box 
parameter-efficient model aligned with the music education community tools, \ie rubrics, which outperforms previous approaches on difficulty estimation. 
In addition, we  
created an interactive companion page for \us{visualizing} CIPI and MKD datasets.
In summary,   
we showed that \us{analyzing} explainable 
descriptors, unlike deep learning models, 
offers clarity, 
which  gives both teachers and students specific insights into pieces.   
This approach not only underscores the importance of explainable artificial intelligence (XAI) in understanding music difficulty, but also \us{emphasizes} the potential for such technologies to contribute to the broader field of music education.
For future research, we consider interesting to creating a dataset based on technical challenges like finger fluency and polyphonic complexity, as well as user studies for understanding the perception of interpretable feedback by music education community.



\section{Ethics Statement}

The system presented in this paper aims at obtaining the difficulty of a musical
piece through several descriptors. In previous work, descriptors were not available, limiting access to the area. This situation underscores the need for open science practices. Therefore, we open our implementation, to facilitate access for new researchers. Besides, the dataset used for this study is available upon request for non-profit and academic research purposes.
While this limits its use in commercial applications, it ensures the reproducibility of the results. The data consists of open-source scores of music that
is no longer copyrighted, its use for open research can thus be considered fair.

The proposed work belongs to the area of assisted music learning. One might
argue that such a tool can have a detrimental impact on music teaching jobs. 
While this is a valid concern, we think that an eventual solution of the addressed task,  would not endanger music educators profession, whose role naturally goes much beyond than  categorizing music in difficulty levels.  Instead, this technology should be seen as a way to support them in the own   teaching practices, for instance, by alleviating their burden on some duties, such as exploring large collections, and by this enabling them to easily discover forgotten  musical works from our cultural heritage which fit students' needs. 
Moreover, through this research, we also aim to convey the message that the path to advancement does not solely lie in acquiring more data or creating larger models. 
By highlighting what drives its decisions, our proposed model aligns with the goals of eXplainable AI, something crucial for its acceptance in music education.  
Although our efforts in making the system interpretable and explainable will partly answer  the common criticisms made to black-box approaches, the real impact of our system remains to be verified by  its future use in real scenarios. 

\vspace{-0.5cm}
\section{Acknowledgements}

We want to thank Alia Morsi's previous work on difficulty estimation from a feature engineering perspective, which encouraged research in tabular data~\cite{morsi2024characterizing}. The team would also acknowledge Marius Miron for continuously insisting on aligning difficulty estimation with explainability. He also highlighted the direction of using a rubric-like explainability feedback, pointing out Ustun et al.'s work~\cite{ustun2019jmlr}. We also thank Roser Batlle-Roca for helping to discuss some concepts between interpretability and explainability based on her research \cite{BatlleRoca2023}.

This work was supported by ``IA y Música: Cátedra en Inteligencia Artificial y Música’' (TSI-100929-2023-1), funded by the Secretaría de Estado de Digitalización e Inteligencia Artificial, and the European Union-Next Generation EU, under the program ``Cátedras ENIA 2022 para la creación de cátedras universidad-empresa en IA’'. This work was also supported by the ANR project TABASCO (ANR-22-CE38-0001) and the travel grant MERMOZ2-012047. 
Finally, this work was also possible through the support of the Hightech Agenda Bayern, funded by the  Free State of Bavaria (Germany).



\bibliography{ISMIRtemplate}

%
%
%
%
%

\end{document}